\documentclass[sigconf]{acmart}
\AtBeginDocument{%
  }

\copyrightyear{2025}
\acmYear{2025}
\setcopyright{cc}
\setcctype{by}
\acmConference[CODASPY '25] {Proceedings of the Fifteenth ACM Conference on Data and Application Security and Privacy}{June 4--6, 2025}{Pittsburgh, PA, USA.}
\acmBooktitle{Proceedings of the Fifteenth ACM Conference on Data and Application Security and Privacy (CODASPY '25), June 4--6, 2025, Pittsburgh, PA, USA}
\acmISBN{979-8-4007-1476-4/25/6}
\acmDOI{10.1145/3714393.3726504}

\usepackage{dirtytalk}
\usepackage{graphicx}
\usepackage{amsmath}
\usepackage{algorithm}
\usepackage{algpseudocode}
\usepackage{subcaption}
\usepackage{multirow}
\usepackage{float}

\patchcmd{\quote}{\rightmargin}{\leftmargin 1.25em \rightmargin}{}{}

\begin{document}

\title[Spend Your Budget Wisely]{Spend Your Budget Wisely: Towards an Intelligent Distribution of the Privacy Budget in Differentially Private Text Rewriting}

\author{Stephen Meisenbacher}
\email{stephen.meisenbacher@tum.de}
\affiliation{%
  \institution{Technical University of Munich \\ School of Computation, Information and Technology}
  \city{Garching}
  \country{Germany}
}

\author{Chaeeun Joy Lee}
\email{chaeeun.joy.lee@tum.de}
\affiliation{%
  \institution{Technical University of Munich \\ School of Computation, Information and Technology}
  \city{Garching}
  \country{Germany}
}

\author{Florian Matthes}
\email{matthes@tum.de}
\affiliation{%
  \institution{Technical University of Munich \\ School of Computation, Information and Technology}
  \city{Garching}
  \country{Germany}
}

\renewcommand{\shortauthors}{Stephen Meisenbacher, Chaeeun Joy Lee, \& Florian Matthes}

\begin{abstract}
The task of \textit{Differentially Private Text Rewriting} is a class of text privatization techniques in which (sensitive) input textual documents are \textit{rewritten} under Differential Privacy (DP) guarantees. The motivation behind such methods is to hide both explicit and implicit identifiers that could be contained in text, while still retaining the semantic meaning of the original text, thus preserving utility. Recent years have seen an uptick in research output in this field, offering a diverse array of word-, sentence-, and document-level DP rewriting methods. Common to these methods is the selection of a privacy budget (i.e., the $\varepsilon$ parameter), which governs the degree to which a text is privatized. One major limitation of previous works, stemming directly from the unique structure of language itself, is the lack of consideration of \textit{where} the privacy budget should be allocated, as not all aspects of language, and therefore text, are equally sensitive or personal. In this work, we are the first to address this shortcoming, asking the question of how a given privacy budget can be intelligently and sensibly distributed amongst a target document. We construct and evaluate a toolkit of linguistics- and NLP-based methods used to allocate a privacy budget to constituent tokens in a text document. In a series of privacy and utility experiments, we empirically demonstrate that given the same privacy budget, intelligent distribution leads to higher privacy levels and more positive trade-offs than a naive distribution of $\varepsilon$. Our work highlights the intricacies of text privatization with DP, and furthermore, it calls for further work on finding more efficient ways to maximize the privatization benefits offered by DP in text rewriting.
\end{abstract}




\maketitle

\section{Introduction}
Efforts to address privacy preservation in Natural Language Processing (NLP) have increased in recent years, notably in the light of rapid advancements in highly advanced AI systems, primarily with Large Language Models (LLMs) \cite{9152761,yan2024protecting}. Such systems have enabled and fostered a nearly unfathomable array of applications for AI and NLP, yet the success of such AI systems is largely contingent upon the large-scale utilization of text data taken from a multitude of data sources, especially the Internet \cite{10198233}. As such, concerns of privacy risks have continued to grow, only exacerbated by the seeming correlation between data usage and model performance \cite{274574,nasr2023scalable}. 

As a response to privacy concerns and vulnerabilities in NLP models, the field of privacy-preserving NLP (PPNLP) has steadily grown in research attention, with methods looking to bolster privacy on the data-, model-, and system-level of NLP applications \cite{lison-etal-2021-anonymisation, sousa2023keep}. A popular choice among researchers in PPNLP is the framework of Differential Privacy (DP), which although was not originally intended for the unstructured domain of text \cite{klymenko-etal-2022-differential}, has seen a number of promising implementations in the literature \cite{hu-etal-2024-differentially}.

Despite the promise, integrating DP into NLP techniques is not simple, and recent literature has unveiled a number of challenges, ranging from loss of semantics and grammatical correctness to difficulties in evaluating privacy preservation in textual data or language models \cite{klymenko-etal-2022-differential, mattern-etal-2022-limits, arnold-etal-2023-guiding}. Beyond this, other works have critiqued the manner in which DP NLP is performed, most notably relaxations in the notion of who is being protected, or what DP guarantee can be provided \cite{vu-etal-2024-granularity, meisenbacher-matthes-2024-thinking}. Looking specifically to the task of \textit{differentially private text rewriting}, in which input texts are rewritten with DP guarantees, it becomes crucial to define on which (syntactic) level a text is rewritten, and what the corresponding guarantee is \cite{vu-etal-2024-granularity}.

\begin{figure}[ht!]
    \centering
    \includegraphics[width=0.99\linewidth]{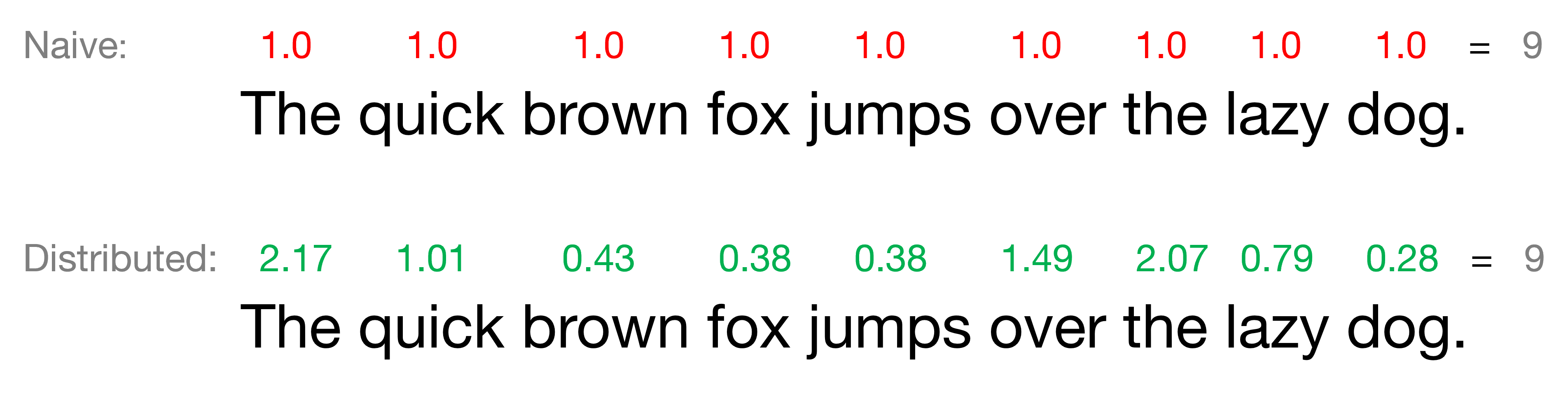}
    \caption{An example of equal privacy budget distribution vs. our proposed approach. Given that certain words may be more sensitive or revealing than others, we propose a more informed distribution (\textit{Distributed}). This example showcases actual output from our proposed toolkit, providing a more sensible budget allocation than equal distribution.}
    \label{fig:dist_example}
\end{figure}

Considering a text \textit{document} as the target quantity to be rewritten, the literature is divided on how a private document can be achieved with DP rewriting. Early techniques considered the \textit{word} to be the unit of privatization, and single-word perturbations could be composed to achieve a document-level guarantee \cite{feyisetan_balle_2020,fernandes2019generalised,yue,Chen2022ACT,carvalho2023tem,10.1145/3643651.3659896}. Later works also looked at the sentence level \cite{Meehan2022SentencelevelPF}, or more conveniently, rewriting an entire document with one DP mechanism run \cite{igamberdiev-habernal-2023-dp,utpala-etal-2023-locally}. While the benefits and limitations of these distinct approaches can be discussed, one unifying limitation is the lack of reasoning in \textit{how} the DP privacy budget (governed by the privacy parameter $\varepsilon$) should be distributed to achieve a privatized document. For example, given word-level DP and a fixed document budget, the naive way may be to divide the overall budget evenly into each word, but this is certainly not optimal (see Figure \ref{fig:dist_example}).

In this work, we propose a more sensible method for DP text rewriting based on one simple thesis: \textit{Not all parts of a document are equally private, and therefore, not all parts of a document should be privatized equally.} Resulting from this, we argue that a method is needed to determine a more intelligent and informed distribution of the privacy budget to a text to be rewritten. Using a toolkit of various linguistics- and NLP-based techniques, we craft a method to distribute a privacy budget sensibly for DP text rewriting, and subsequently, we leverage compositionality to achieve a final privatized text which fits into the constraints of the budget. In doing so, we answer the following research question in this work:

\begin{quote}
    \textit{How can one intelligently \say{distribute} a given privacy budget in differentially private text rewriting, and what is the resulting effect on the utility and privacy of the privatized data?}
\end{quote}

To test the utility- and privacy-preservation of our method, we compare the downstream task performance and resistance to adversarial attacks of privatized data using our distribution method to data which is \textit{naively} privatized. We find that distributing the privacy budget with our proposed toolkit generally increases the privacy of DP rewritten text, while also leading to better trade-offs in certain cases. On the other hand, privacy budget distribution nearly always leads to lower utility and lower text coherence, leading us to critically analyze the merits and limitations of our toolkit.

As a result of our work and based on our empirical findings, we make the following contributions to the DP NLP field:
\begin{enumerate}
    \item We are the first to consider the distribution of privacy budget for DP text rewriting, and we propose a toolkit to determine sensible budget allocations given an input text. The toolkit is available at \url{https://github.com/sjmeis/EpsilonDistributor}.
    \item We evaluate our method in a series of privacy and utility evaluations, showing the effectiveness of budget distribution in privacy preservation.
    \item We critically analyze and discuss the implications of intelligent budget distribution for DP text rewriting, proposing ways forward to build upon our work.
\end{enumerate}

\section{Foundations}
\label{sec:foundations}
\subsection{Differential Privacy}
Formalized nearly two decades ago, Differential Privacy (DP) \cite{dwork2006differential} guarantees that any computation performed on a database, or a more general collection of databases, is nearly the same regardless of the inclusion or exclusion of a single data point. Formally, given two databases $\mathcal{D}$ and $\mathcal{D^\prime}$ differing in only one data point, any query or computation run on $\mathcal{D}$ and $\mathcal{D^\prime}$ will yield \textit{similar} results when utilizing some DP mechanism $\mathcal{M}$. Such databases that differ only by a single element are called \textit{neighboring} or \textit{adjacent} databases.

\begin{definition}
\label{def:DP}
$(\varepsilon,\delta)$-Differential Privacy.
A mechanism $\mathcal{M}: \mathcal{X}^m \to \mathcal{O}$ operating on any two adjacent databases $\mathcal{D}$, $\mathcal{D^\prime} \in \mathcal{X}^m$ is $(\varepsilon,\delta)$-differentially private, iff $\forall O \subseteq \mathcal{O}$, the following holds:
\begin{displaymath}
    \mathbb{P}[\mathcal{M}(\mathcal{D}) \in O] 
     \leq e^\varepsilon \cdot {\mathbb{P}[\mathcal{M}(\mathcal{D^\prime}) \in O]} + \delta
\end{displaymath}
where $\varepsilon > 0$ and $\delta \in [0, 1]$
\end{definition}

Intuitively, ensuring the above privacy guarantee grants plausible deniability to individuals participating in a database, such that the result of some query cannot be attributed to this person's participation in a database. Instead, the DP mechanism $\mathcal{M}$ grants this deniability, usually achieved by the injection of calibrated random noise to queries or computations. 

In our work, we utilize the \textsc{DP-BART} rewriting mechanism \cite{igamberdiev-habernal-2023-dp}, which guarantees $(\varepsilon,\delta)$-DP for any two \textit{documents}.

\subsection{Metric Differential Privacy}
In some domains, such as that of natural language and textual data, the original notion $(\varepsilon,\delta)$-DP may be too restrictive, or rather not fitting to the reasoning of the \say{individual} in a dataset. As such, the notion of \textit{Metric} Differential Privacy (MDP) has emerged in recent years to address the limitation \cite{chatzikokolakis2013broadening}. It is most useful when dealing in \textit{metric spaces}, and it can be defined as follows.

Let $\mathcal{X}$ and $\mathcal{Z}$ be finite sets and let $d: \mathcal{X} \times \mathcal{X} \to \mathbb{R+}$ be a distance metric defined on the set $\mathcal{X}$.

\begin{definition}
($d_\mathcal{X}$-privacy). Let $\varepsilon > 0$. A randomized mechanism $\mathcal{M}: \mathcal{X} \to \mathcal{Z}$ satisfies $\varepsilon d_\mathcal{X}$-privacy iff $\forall x, x^\prime \in \mathcal{X}$ and $\forall z \in \mathcal{Z}$

	\begin{equation}
		\frac{ \mathbb{P}[ \mathcal{M}(x) = z ] }
			 { \mathbb{P}[ \mathcal{M}(x^\prime) = z ] } 
		\leq e^{\varepsilon d(x, x^\prime)}
		\label{eq: edp}
	\end{equation}
\end{definition}

The above can clearly be seen as a relaxation of Definition \ref{def:DP}, as the privacy guarantee is now scaled according to the distance between any two data points in a given space. Intuitively, when an MDP mechanism is applied, queries on data points which are close in space, as measured by the chosen metric, would yield more \say{similar} output distributions as compared to points farther apart.

In this work, we utilize the \textsc{1-Diffractor} mechanism \cite{10.1145/3643651.3659896}, which leverages word-level MDP to provide guarantees for any two \textit{words}.

\subsection{Local DP and Text Rewriting}
As opposed to the setup where data is first collected by some aggregator before applying a chosen DP mechanism, known as \textit{Global} Differential Privacy, the concept of \textit{Local} Differential Privacy (LDP) becomes useful in cases where third party aggregators are not trusted or where privatization must occur on the user side. In LDP, the notion of adjacent databases is shifted to the individual, and it is defined over data points from a single individual. Thus, every collected data point from a single individual is adjacent to every other data point from another individual \cite{4690986}. Note that LDP is also defined for MDP, thus yielding MLDP \cite{meisenbacher-etal-2024-comparative}.

In the case of DP text rewriting, the LDP setup is most sensible, so that users can privatize their text(s) via rewriting before sharing it with third parties. In this scenario, the user utilizes a DP mechanism operating on a particular syntactic level, and they privatize their textual data accordingly. For example, in a word-level scenario, a user shares each obfuscated word, whereby documents can be shared according to the composition theorem of DP (see next). Similarly, if the users opts to use a document-level mechanism, the output of each mechanism run is a privatized document with an accompanying privacy guarantee. As noted by \citet{vu-etal-2024-granularity}, the distinction of granularity is particularly crucial in the case of text privatization with DP, as it must be made transparent for which syntactic unit a guarantee is being provided.

The limitation with the LDP setup, however, is that for the given unit of protection (e.g., a word or document), any data point from one user is adjacent to the entire space of data points. For example, in the case of documents, any potential text document is adjacent to any other document. This limitation is highlighted by \citet{igamberdiev-habernal-2023-dp}, which above all necessitates higher privacy budgets for sensible privatization. 

\subsection{Composition in DP Text Rewriting}
When reporting privacy guarantees in DP text rewriting scenarios, it becomes very important to leverage the composition theorem of DP, which is defined as follows:

\begin{theorem}
    Composition in DP \cite{dwork2006differential}. 
    \\ Let $M_1$ be an $\varepsilon_1$-differentially private algorithm, and let $M_2$ be an $\varepsilon_2$-differentially private algorithm. Then their combination, defined to be $M_{1,2}$: $M_{1,2}$(x) = ($M_1$(x), $M_2$(x)), is ($\varepsilon_1 + \varepsilon_2$)-differentially private.
\end{theorem}

The implications of the above theorem are quite useful in reporting aggregate privacy guarantees: if one runs a DP mechanism with privacy budget $\varepsilon$ for $n$ times, then the resulting guarantee is $n \cdot \varepsilon$. The intuition is also clear; privacy guarantees begin to degrade the more times a mechanism is used on the same data.

In DP text rewriting, composition can be leveraged to utilize DP mechanisms a number of times to achieve the desired syntactic unit of privatization. For example, given a text of 10 words, a word-level DP mechanism can be run on each of the 10 words with a budget of $\varepsilon$ per word for a total guarantee of $10\varepsilon$. This can naturally be extrapolated to sentences in a document, documents per user, and beyond. For the purposes of this work, we treat the document as the final unit of protection, although recent work has shown that this assumption may not always be correct \cite{vu-etal-2024-granularity}.

In this work, we place a particular focus on the question of composition in DP text rewriting, investigating whether this theorem can be leveraged more wisely by considering the hierarchical nature of textual data. Specifically, we consider the scenario where a fixed privacy budget is allotted for each document to be privatized, and we explore how this budget can be maximized to protect the privacy concealed in natural language, while still maintaining the utility of text datasets. We challenge the \say{naive} distribution of privacy budgets, in which, for example, a document is privatized singly without higher focus on more sensitive sentences, or similarly where a document is privatized with equal emphasis on all words rather than an intelligent distribution of stricter privatization to more sensitive words. To address this, we now introduce a toolkit of techniques that will allow for a more informed and sensible privacy budget distribution in DP text rewriting.

\section{A Budget Distribution Toolkit for DP Text Rewriting}
In this section, we introduce the underlying methodology behind the distribution of a given privacy budget over a text document. The goal of such a distribution is to allocate a privacy budget intelligently so as to account for a number of linguistic- and NLP-informed factors which may make certain tokens in a document more sensitive than others. We first outline the general framework for budget distribution, and then we proceed to introduce the individual components of our proposed toolkit.

\subsection{Allocating $\varepsilon$}
\label{sec:allocation}
We consider an arbitrary text document $\mathcal{D}$ to be privatized via DP text rewriting. The document $\mathcal{D}$ consists of $n$ tokens, or words, which are sequential in nature, i.e., $D = (t_i)^1_n$, where $t_i$ denotes a token in the $i$-th position in the document string.

We also define a general scoring function $\mathcal{S(\mathcal{D}, \varepsilon)}$, which takes as input an arbitrary document $\mathcal{D}$ and a privacy budget $\varepsilon$. The output of $\mathcal{S}$ is a mapping $\mathcal{S}: t_i \rightarrow \mathbb{R}^+ = s_i, \forall i \in \{1..n\}$. Thus, each token in the document $\mathcal{D}$ is assigned a normalized \textit{sensitivity score} $s$, where a higher score denotes a greater \say{need} for privatization. In the context of DP and the total privacy budget $\varepsilon$, this translates to the allocation of a smaller token budget. Formally, we are therefore solving the linear equation $Ax = b$, where

\begin{displaymath}
    A = \begin{bmatrix} \frac{1}{s_1} & \frac{1}{s_2} & ... & \frac{1}{s_{n-1}} & \frac{1}{s_n}\end{bmatrix}, b = \varepsilon
\end{displaymath}
\\
Thus, the resulting solution for the equation $\textbf{x}$ is a 1:1 mapping of constituent tokens to per-token budget allocations:

\begin{displaymath}
    \textbf{x} = \begin{bmatrix} x_1 & x_2 & ... & x_{n-1} & x_n \end{bmatrix} = \begin{bmatrix} \varepsilon_1 & \varepsilon_2 & ... & \varepsilon_{n-1} & \varepsilon_n \end{bmatrix}
\end{displaymath}
\\
And finally, by leveraging compositionality, we can achieve a distribution that respects the original total privacy budget $\varepsilon$:

\begin{displaymath}
    \sum^n_{i=1} \varepsilon_i = \varepsilon
\end{displaymath}

Note that the case of per-token privacy budget allocations can be generalized to the sentence level in document privatization by simply summing scores of the constituent tokens of a sentence, thus yielding a sentence-level privacy budget allocation.

\subsection{Budget Allocation Methods}
Our toolkit consists of five methods used to calculate per-token budget allocations, which are introduced below, as well as the technique used to combine the component scores into a final allocation. For full method details, we refer to reader to our code repository. 

\subsubsection{Information Content (\textbf{IC})}
Information Content (IC), also referred to as \textit{self-information} or \textit{Shannon information}, is a value derived from measuring the probability of a particular event occurring. In the context of linguistics, such a value can be assigned to any given unit in language, most notably a word token.

In leveraging IC measures, we base the resulting scores on the hypothesis that the greater the relative information given by a particular word is, the greater the need for privacy. We utilize the IC measures provided by the \textsc{nltk} packages, namely the \textit{semcor}, \textit{brown}, \textit{bnc}, \textit{shaks}, and \textit{treebank} corpora. We also use the English WordNet, which contains \textit{synsets}; these synsets, or word entries, are required as the input format for retrieving the IC scores.

As the abovementioned corpora only assign IC scores to nouns and verbs, we likewise only score nouns and verbs; these IC scores are given on the positive integer range (e.g., \textit{IC}(`dog') = 235). Non-noun/verb tokens, as well as tokens not existing in the corpora, are assigned a score of 1, or the lowest possible positive integer. 

\subsubsection{Part-of-Speech Informativeness (\textbf{POS})}
Continuing with our hypothesis that word informativeness can help to determine the level of privacy needed, we also leverage Part-of-Speech (POS) information to calculate informativeness scores. POS tags indicate the grammatical function of a word (noun, verb, adjective, etc.) in a text. Therefore, assigning different weights to different POS tags can reflect their typical importance in conveying meaning. For example, verbs often play a more central role in a sentence compared to prepositions. This helps distinguish between function words (like articles and prepositions) and content words (nouns, verbs, adjectives) that contribute more to the core meaning \cite{klammer2010analyzing}.

To define a weighting scheme for different POS tags, we refer to a previous study based on Twitter data \cite{gimpel-etal-2011-part}, and use the aggregate statistics to derive weights that denote the relative frequency of POS tags. We focus on \textit{nouns} (NN), \textit{pronouns} (PR), \textit{verbs} (VB), \textit{adjectives} (JJ), \textit{adverbs} (RB), and \textit{numbers} (CD), with the following weights: \{NN: 14, PR:7, VB:15, CD:2, JJ:5, RB:5\}.
All POS tags outside this set are not considered sensitive, and they receive a weight of 0.1, chosen to be distinct from the abovementioned values yet to assign non-zero weights to avoid division errors.

\subsubsection{Named Entity Recognition (\textbf{NER})}
The task of Named Entity Recognition (NER) aims to identify \textit{named entities} in a given text, such as names, locations, and organizations. These entities are typically very important to a particular sentence's meaning; however, they are generally quite identifying, such as with names. 

We use the NER tool provided by the \textsc{spaCy} package\footnote{\url{https://spacy.io/}} to identify named entities in a given input document, and all tokens that belong to a named entity are assigned a score of 1, otherwise a score of 0.

\subsubsection{Word Importance (\textbf{WI})}
This method compares the semantic similarity between the entire text document and each individual word. Words with a larger difference in similarity likely contribute more to the overall meaning, as they introduce new semantic meaning not already conveyed by the rest of the text.

To measure such importance of each word, we iteratively remove each word token from a given text, and measure its similarity to the remainder of the text. The lower the similarity between these two entities, the greater the importance of the word. 

\subsubsection{Sentence Difference (\textbf{SD})}
In a similar way to the above \textit{Word Importance} scoring, we also measure the semantic difference between a given text and the same text without a single word. This, similar to the above, provides a notion of a word's importance semantically. Words whose removal causes a more significant drop in similarity are considered more important because their absence significantly impacts the overall meaning. Thus, such words are more identifiable in text and must be treated with higher privacy.

To measure \textit{Sentence Difference}, we create $n$ versions of the original sentence, each with one of $t_1...t_n$ removed. These versions are then compared semantically to the original, unaltered sentence, and the resulting scores are assigned to corresponding tokens. 

For both the \textbf{WI} and \textbf{SD} methods, we utilize the \textsc{thenlper/gte-small} embedding model\footnote{\url{https://huggingface.co/thenlper/gte-small}} \cite{li2023generaltextembeddingsmultistage}.

\subsection{Calculating a Final Budget Distribution}
Given the scores outputted by each of the five methods described above, the final steps involve combining these scores for a final privacy budget distribution for DP text rewriting.

First, all score sets, which map a score to each token in an input document, are normalized between 0 and 1. This is particularly necessary in the case of \textbf{POS}, where the weights assigned do not fall in the range [0,1]. Then, the average score of each token amongst the five scoring methods is taken to achieve an aggregate score for each token. Note that we assume an equal weighting for each of the methods, and we do not experiment with different weighted averages. However, please refer to Section \ref{sec:ablation} and Table \ref{tab:ablation} for the results of our ablation study for the described methods. 

Given the aggregate scores for all word tokens in a document, the linear equation as described in Section \ref{sec:allocation} is solved. This results in an individual budget for each token, all of which add up to the total allocated privacy budget $\varepsilon$.

\section{Experimental Setup and Results}
Following the guidelines of \citet{mattern-etal-2022-limits} of what comprises an effective text privatization method, we evaluate the performance of our proposed budget distribution method on two primary categories: privacy protection and utility preservation. These evaluations will demonstrate to what degree our method improves upon the empirical privacy protections afforded by DP text rewriting, while simultaneously testing whether utility in downstream tasks and text coherence can still be achieved. In the following, we outline the full methodological design of our experiments, as well as provide the corresponding results.

\subsection{DP Text Rewriting Methods}
In the scope of this work, we choose two DP text rewriting methods from the recent literature, which will serve as the testbed for our proposed budget distribution toolkit.

\paragraph{\textsc{1-Diffractor} \cite{10.1145/3643651.3659896}}
\textsc{1-Diffractor} is a word-level MLDP text obfuscation mechanism proposed by \citeauthor{10.1145/3643651.3659896} to improve the efficiency of previous word-level mechanisms. In essence, the mechanism \textit{perturbs} words in a document by adding DP noise to word embeddings in one-dimensionally sorted lists. In this work, we use the \textit{geometric} version of the \textsc{1-Diffractor} mechanism, or $1-D_G$. Following from the original work, we test the following base, per-word $\varepsilon$ values: $\varepsilon \in \{0.1, 0.5, 1.0\}$.

\paragraph{\textsc{DP-BART} \cite{igamberdiev-habernal-2023-dp}}
\textsc{DP-BART} is a DP text rewriting mechanism proposed by \citeauthor{igamberdiev-habernal-2023-dp} which achieves DP rewriting at the document level by adding calibrated DP noise in the latent space representation of the \textsc{BART} encoder-decoder model \cite{lewis-etal-2020-bart}. In this work, we utilize the \textsc{DP-BART-CLV} version, which achieves DP by clipping the latent space before adding noise. Following from the original work, we choose the clipping range of $[-0.1, 0.1]$, and we test on the following document-level $\varepsilon$ values: $\varepsilon \in \{500, 1000, 1500\}$.

\subsubsection{Setting the total privacy budget}
In the evaluation of both of the abovementioned rewriting methods on our chosen datasets and tasks (see next), we must first determine the total privacy budget ($\varepsilon$) allocated to each text document to be privatized. This is especially pertinent in the case of our chosen \textit{word-level} mechanism, as we aim to privatize documents at the \textit{document} level.

Following the example set in previous work \cite{meisenbacher-etal-2024-collocation}, we fix a \textit{dataset-specific} per-document privacy budget, which can be derived as the chosen base $\varepsilon$ value, scaled (multiplied) by the average number of tokens in a document for a given dataset. Thus, for \textsc{1-Diffractor}, our chosen values of $\varepsilon \in \{0.1, 0.5, 1.0\}$ will be scaled for each dataset to achieve the total privacy budget available for each document. The exact values used for these calculations and the per-document budgets are made apparent in all tables presenting results.

Note that for \textsc{DP-BART}, the chosen $\varepsilon$ values already represent the per-document budgets, as this mechanism operates directly on the document level. For both \textsc{1-Diffractor} and \textsc{DP-BART}, we test for both an equal distribution of the total privacy budget, i.e., $\varepsilon / num\_tokens$, as well as budget distribution with our proposed toolkit. With \textsc{DP-BART}, the resulting per-word budgets are summed to achieve sentence-level budgets. Thus, an input document is split into its component sentences, and each sentence is privatized with \textsc{DP-BART} according to the allocated budget. In all cases, stopwords (common English words as determined by the \textsc{nltk} package) are not considered in the budget distribution and are not privatized.

\subsection{Privacy Experiments}
We run experiments to evaluate the privacy-preserving capability of our method relative to naive (equal) budget distribution. These experiments take two forms: empirical privacy and membership inference. We first introduce the datasets used for experimentation, and then we proceed to describe in detail the evaluation procedures.

\subsubsection{Datasets and Tasks}
For the privacy experiments, we leverage three existing public datasets.

\paragraph{Yelp Reviews}
We utilize a dataset of reviews from the \textit{Yelp} platform, specifically the subset made available by \citet{utpala-etal-2023-locally}\footnote{The full dataset is available at \url{https://huggingface.co/datasets/Yelp/yelp_review_full}.}. This subset contains 17,295 reviews from 10 distinct users on the platform. Each review is denoted as a \textit{positive} or \textit{negative} review in terms of sentiment. This dataset's makeup allows for an adversarial \textit{authorship identification} task, in which an attacker's goal is to guess the identity of the text's author given only the text.

\paragraph{Trustpilot Reviews}
Made available by \citet{10.1145/2736277.2741141}, the Trustpilot Reviews corpus is a collection of reviews in several languages from the Trustpilot platform. We only use English-language reviews from the \textit{en-US} subset, and we take a 10\% sample (29,490 reviews). Each review is marked as positive or negative, as well as with the gender (male/female) of the author, creating the opportunity for evaluation on an adversarial \textit{gender identification} task.

\paragraph{Blog Corpus}
The final dataset we use for our privacy experiments is a subset from the Blog Authorship Corpus \cite{schler2006effects}, a large collection of user-written blog posts on an internet forum. In particular, we make use of the \textit{author10} split made available by \citet{meisenbacher-matthes-2024-thinking}, which contains a total of 15,070 blog posts from the top-10 contributing authors. Thus, we create another authorship identification scenario for our empirical privacy experiments.

\begin{table*}
  \centering
    \caption{Empirical Privacy Results. Adversarial scores are represented by micro-F1 scores, where \textit{(s)} denotes the \textit{static} attacker and \textit{(a)} denotes the \textit{adaptive} attacker. $\gamma$ refers to the \textit{Relative Gain} for both the static and adaptive settings. For all \textsc{1-Diffractor} tasks, we report the per-word $\varepsilon$, as well as the total allocated budget (indicated in parentheses), which is calculated by (\textit{per-word $\varepsilon$})$\cdot$(\textit{Avg. Tokens}), or average number of tokens in a document per dataset. Baseline scores, i.e., adversarial performance on the non-privatized data, are also provided. Where $\gamma$ scores are reported, the \textbf{bolded} score represents the better pairwise score between a ``naive'' distribution and our proposed method.}
  \begin{subtable}{\textwidth}
    \centering
       \resizebox{\textwidth}{!}{
    \begin{tabular}{lc|cc|cc|cc||cc|cc|cc}
      \multicolumn{2}{c|}{\textbf{Yelp}} & \multicolumn{6}{c||}{\textbf{1-Diffractor}} & \multicolumn{6}{c}{\textbf{DP-BART}} \\ \hline
    \multicolumn{2}{r|}{Avg. Tokens} & \multicolumn{6}{c||}{181.06} &  \multicolumn{6}{c}{181.06}  \\ \hline
      \multicolumn{2}{r|}{$\varepsilon$} & \multicolumn{2}{c|}{0.1 (18.11)} & \multicolumn{2}{c|}{0.5 (90.53)} & \multicolumn{2}{c||}{1 (181.06)} & \multicolumn{2}{c|}{500} & \multicolumn{2}{c|}{1000} & \multicolumn{2}{c}{1500} \\ \hline
      \multicolumn{2}{l|}{Distribution Method} & Naive & Ours & Naive & Ours & Naive & Ours & Naive & Ours & Naive & Ours & Naive & Ours \\ \hline
      \multicolumn{1}{l|}{Utility (F1) $\uparrow$}  & $95.09_{0.3}$ & $93.53_{0.0}$ & $93.53_{0.0}$  & $95.01_{0.6}$  & $94.01_{0.9}$  & $94.70_{1.0}$ & $94.45_{0.6}$  &  $93.53_{0.0}$ & $93.53_{0.0}$  & $93.99_{0.1}$  &  $93.53_{0.0}$ & $94.47_{0.7}$ &  $93.53_{0.0}$ \\
      \multicolumn{1}{l|}{PP+ $\uparrow$}         & -156  & -312  &  -312 &  -164 &  -264 &  -195 &  -220 & -312  & -312  &  -266 &  -312 &  -218 & -312  \\ \hline
      \multicolumn{1}{l|}{Adv. F1 (s) $\downarrow$}  &  95.90 & 42.20 & 42.37 & 57.23 & 55.09 & 65.84 & 62.60 & 26.24 & 17.11 & 26.99 & 15.09 & 27.86 & 15.55 \\
      \multicolumn{1}{l|}{Adv. F1 (a) $\downarrow$}  & 95.90  & $80.92_{3.0}$ &  $82.35_{0.9}$ & $92.16_{0.1}$ & $88.44_{1.4}$  & $92.72_{0.5}$  & $92.87_{0.7}$  & $38.82_{1.0}$  & $32.87_{0.4}$ &  $61.21_{0.6}$ & $37.30_{0.8}$ & $67.63_{1.2}$ &  $38.71_{0.6}$ \\ \hline
      \multicolumn{1}{l|}{$\gamma$ (s) $\uparrow$} & -  & \textbf{1.81} & \textbf{1.81} & 0.64 & \textbf{1.31} & 0.19 & \textbf{0.91} & 2.06 & \textbf{2.19} & 1.75 & \textbf{2.22} & 1.43 & \textbf{2.22} \\
      \multicolumn{1}{l|}{$\gamma$ (a) $\uparrow$} & -  & \textbf{1.23} & 1.21 & 0.11 & \textbf{0.81} & 0.30 & \textbf{0.46} & 1.86 & \textbf{1.95} & 1.23 & \textbf{1.89} & 0.83 & \textbf{1.87}
    \end{tabular}
    }
    \caption{Yelp}
  \end{subtable}
  \begin{subtable}{\textwidth}
    \centering
       \resizebox{\textwidth}{!}{
    \begin{tabular}{lc|cc|cc|cc||cc|cc|cc}
      \multicolumn{2}{c|}{\textbf{Trustpilot}} & \multicolumn{6}{c||}{\textbf{1-Diffractor}} & \multicolumn{6}{c}{\textbf{DP-BART}} \\ \hline
      \multicolumn{2}{r|}{Avg. Tokens} & \multicolumn{6}{c||}{51.23} &  \multicolumn{6}{c}{51.23}  \\ \hline
      \multicolumn{2}{r|}{$\varepsilon$} & \multicolumn{2}{c|}{0.1 (5.12)} & \multicolumn{2}{c|}{0.5 (25.62)} & \multicolumn{2}{c||}{1 (51.23)} & \multicolumn{2}{c|}{500} & \multicolumn{2}{c|}{1000} & \multicolumn{2}{c}{1500} \\ \hline
      \multicolumn{2}{l|}{Distribution Method} & Naive & Ours & Naive & Ours & Naive & Ours & Naive & Ours & Naive & Ours & Naive & Ours \\ \hline
      \multicolumn{1}{l|}{Utility (F1) $\uparrow$} &  $99.49_{0.1}$ &  $94.87_{1.3}$ &  $93.62_{2.3}$ &  $98.15_{0.2}$ & $97.93_{0.3}$ & $98.89_{0.1}$ & $98.64_{0.6}$  & $92.59_{0.4}$  & $92.09_{0.1}$  &  $98.03_{0.1}$ & $92.16_{0.2}$  & $98.51_{0.1}$ & $93.36_{0.1}$  \\
      \multicolumn{1}{l|}{PP+ $\uparrow$}          & 366  & -96  & -221  & 232  &  210 &  306 &  281 &  -324 & -374  &  220 &  -367 & 268  & -247  \\ \hline
      \multicolumn{1}{l|}{Adv. F1 (s) $\downarrow$}  & 72.16 & 59.61 & 59.51 & 64.12 & 63.45 & 67.28 & 66.40 & 60.71  & 58.94 & 59.85 & 59.34 & 60.16 & 59.34 \\
      \multicolumn{1}{l|}{Adv. F1 (a) $\downarrow$}  & 72.16 & $60.71_{3.7}$ & $60.37_{3.2}$ & $67.70_{2.2}$ & $66.59_{0.6}$ & $63.87_{4.1}$ & $68.25_{2.2}$ & $58.83_{0.6}$ & $58.09_{0.0}$ & $62.48_{0.9}$ &  $58.12_{0.0}$ & $61.18_{1.8}$ & $58.94_{1.2}$ \\ \hline
      \multicolumn{1}{l|}{$\gamma$ (s) $\uparrow$} & -  & \textbf{1.02} & 0.70 & 1.09 & \textbf{1.16} & \textbf{1.34} & 0.81 & 0.20 & \textbf{0.38} & \textbf{1.84} & 0.33 & \textbf{1.91} & 0.66 \\
      \multicolumn{1}{l|}{$\gamma$ (a) $\uparrow$} & -  & \textbf{0.82} & 0.54 & 0.44 & \textbf{0.59} & \textbf{1.34} & 0.48 & \textbf{0.54} & \textbf{0.54} & \textbf{1.36} & 0.55 & \textbf{1.73} & 0.73
    \end{tabular}
    }
    \caption{Trustpilot}
  \end{subtable}
  \begin{subtable}{\textwidth}
    \centering
    \resizebox{\textwidth}{!}{
    \begin{tabular}{lc|cc|cc|cc||cc|cc|cc}
      \multicolumn{2}{c|}{\textbf{Blog}} & \multicolumn{6}{c||}{\textbf{1-Diffractor}} & \multicolumn{6}{c}{\textbf{DP-BART}} \\ \hline
      \multicolumn{2}{r|}{Avg. Tokens} & \multicolumn{6}{c||}{53.94} &  \multicolumn{6}{c}{53.94}  \\ \hline
      \multicolumn{2}{r|}{$\varepsilon$} & \multicolumn{2}{c|}{0.1 (5.39)} & \multicolumn{2}{c|}{0.5 (26.97)} & \multicolumn{2}{c||}{1 (53.94)} & \multicolumn{2}{c|}{500} & \multicolumn{2}{c|}{1000} & \multicolumn{2}{c}{1500} \\ \hline
      \multicolumn{2}{l|}{Distribution Method} & Naive & Ours & Naive & Ours & Naive & Ours & Naive & Ours & Naive & Ours & Naive & Ours \\ \hline
      \multicolumn{1}{l|}{Adv. F1 (s) $\downarrow$} & 58.64  & 26.56 & 26.19 & 31.34 & 31.96 & 35.39 & 35.43 & 12.73 & 7.95 & 15.05 & 12.51 & 15.87 & 17.22 \\
      \multicolumn{1}{l|}{Adv. F1 (a) $\downarrow$} &  58.64 & $40.05_{0.8}$ & $38.55_{1.0}$  & $44.89_{1.7}$ &  $44.49_{0.8}$ & $43.96_{9.6}$ & $46.89_{1.5}$ & $13.61_{0.5}$ & $8.95_{0.8}$ & $21.77_{0.9}$ & $13.58_{1.9}$ &  $23.77_{1.0}$ & $16.60_{3.5}$
    \end{tabular}
    }
    \caption{Blog}
  \end{subtable}
  \label{tab:ep_results}
\end{table*}

\subsubsection{Empirical Privacy Evaluation}
\label{sec:ep_eval}
The first of two overarching privacy evaluation tasks takes the form of \textit{empirical privacy} evaluations. Here, we test the ability of DP text rewriting to reduce the adversarial advantage (i.e., attribute inference performance) on authorship or gender identification, measured \textit{empirically}.

To test empirical privacy, we first privatize all of the above datasets using our two chosen DP rewriting methods under the chosen privacy budgets. This is done for both naive budget distribution and distribution using our toolkit. Then, we train an adversarial classification model to predict the protected attribute (author or gender) given a text. For all experiments, a \textsc{deberta-v3-base} model \cite{he2021deberta} is used, and datasets are split into a 90\% train / 10\% test set.

We perform the adversarial training in two settings, following the recent literature \cite{mattern-etal-2022-limits, utpala-etal-2023-locally, meisenbacher-matthes-2024-thinking}. In the first, called the \textit{static} setting, the adversarial classification model is trained on the non-privatized train set, and the resulting model is evaluated on the privatized test set for the \textit{static} results. This models a less capable attacker who does not have knowledge of the DP rewriting method. In the more capable setting, called the \textit{adaptive} attacker, the adversarial model is trained on the privatized train set, and then evaluated on the privatized test set, thus mimicking an adversary who is able to train a better model given the ability to align the training dataset. For all scenarios in these experiments and for the remainder of this work, training is performed for one full epoch, with a batch size of 32, maximum input length of 512 tokens, learning rate of 1e-5, and otherwise default HuggingFace \textsc{Trainer} parameters. All training procedures are repeated three times on different random shuffles of the train set, and the report scores represent the average score with standard deviation. The hardware used is an RTX A6000 GPU.

For score reporting, we first measure the corresponding utility of each privatized dataset, measured by the micro-F1 score on the binary sentiment analysis task after one epoch of training. As the datasets are imbalanced (many positive reviews), we also provide \textit{PP+}, or the percentage points achieved over majority-class guessing.

Next, we report the adversarial F1 score against the non-privatized (plaintext) baseline, where a lower score denotes that the DP rewriting has better protected privacy. Finally, we report the \textit{Relative Gain ($\gamma$)} metric \cite{10.1145/3485447.3512232, mattern-etal-2022-limits}, which aims to illustrate the balance between (potential) utility lost and privacy gained. Let $P_o$, $U_o$ represent the baseline privacy and utility scores, respectively, and $P_r$, $U_r$ be the scores observed on the privatized datasets. Relative Gain is thus defined as $\gamma = (U_r / U_o) - (P_r / P_o)$, with the higher the better. Different to previous work, we calculate the change in F1 over random / majority-class guessing on the validation set, denoted $MG_u$ (utility) and $MG_p$ (privacy), as the Trustpilot and Yelp datasets are imbalanced; thus $RG = \frac{U_r - MG_u}{U_o - MG_u} - \frac{P_r - MG_p}{P_o - MG_p}$.

In the Yelp dataset, the 10\% validation split contains 1618 positive reviews and 112 negative reviews. Thus, $MG_u = 96.65$. The split contains 304 reviews from the most frequent author, with the nine other authors writing 1426 reviews. Thus, $MG_p = 29.89$, showing the majority-class guessing performance. In the Trustpilot dataset, the 10\% validation split contains 2713 positive reviews and 236 negative reviews. Thus, $MG_u = 95.83$. The split contains 1713 reviews from males and 1236 reviews from females. Thus $MG_p = 66.67$. Note that we use random guessing performance due to the relative balance between male and female authors.

The complete results of the empirical privacy experiments can be found in Table \ref{tab:ep_results}, for both the static (s) and adaptive (a) settings. Note that as the Blog dataset does not have an associated utility task, we do not report utility or $\gamma$ values.

\begin{table}[htbp]
  \centering
    \caption{Membership Inference Evalaution. \textbf{Bolded} scores represent the better score between ``naive'' and our proposed distribution (shown only for $MTI_{bow}$ and NN). For further details on each metric, please refer to Section \ref{sec:mi}.}
  \begin{subtable}{\linewidth}
    \centering
    \resizebox{\linewidth}{!}{
   \begin{tabular}{lcl|c|c|c|c|c}
 &  &  & $MTI_{seq}T1\downarrow$ & $MTI_{seq}T3\downarrow$  & $MTI_{bow}T1\downarrow$ & $MTI_{bow}T3\downarrow$ & NN$\uparrow$ \\ \hline
\multicolumn{1}{l|}{} & \multicolumn{1}{c|}{0.1} & Naive & 0.003 & 0.007 & \textbf{0.121} & \textbf{0.151} & 111 \\
\multicolumn{1}{l|}{} & \multicolumn{1}{l|}{} & Ours & 0.003 & 0.007 & \textbf{0.121} & \textbf{0.151} & \textbf{154} \\ \cline{2-8} 
\multicolumn{1}{l|}{Yelp} & \multicolumn{1}{l|}{0.5} & Naive & 0.003 & 0.006 & 0.122 & \textbf{0.153} & 9 \\
\multicolumn{1}{l|}{} & \multicolumn{1}{l|}{} & Ours & 0.003 & 0.006 & \textbf{0.120} & \textbf{0.153} & \textbf{17} \\ \cline{2-8} 
\multicolumn{1}{l|}{} & \multicolumn{1}{l|}{1} & Naive & 0.003 & 0.006 & 0.124 & 0.155 & 2 \\
\multicolumn{1}{l|}{} & \multicolumn{1}{l|}{} & Ours & 0.003 & 0.006 & \textbf{0.123} & \textbf{0.154} & \textbf{4} \\ \hline
\multicolumn{1}{l|}{} & \multicolumn{1}{l|}{0.1} & Naive & 0.002 & 0.004 & \textbf{0.055} & \textbf{0.074} & 308 \\
\multicolumn{1}{l|}{} & \multicolumn{1}{l|}{} & Ours & 0.002 & 0.004 & \textbf{0.055} & 0.075 & \textbf{383} \\ \cline{2-8} 
\multicolumn{1}{l|}{Trustpilot} & \multicolumn{1}{l|}{0.5} & Naive & 0.002 & 0.004 & 0.059 & 0.078 & 28 \\
\multicolumn{1}{l|}{} & \multicolumn{1}{l|}{} & Ours & 0.002 & 0.004 & \textbf{0.057} &\textbf{ 0.077} & \textbf{64} \\ \cline{2-8} 
\multicolumn{1}{l|}{} & \multicolumn{1}{l|}{1} & Naive & 0.002 & 0.004 & 0.062 & 0.081 & 5 \\
\multicolumn{1}{l|}{} & \multicolumn{1}{l|}{} & Ours & 0.002 & 0.003 & \textbf{0.060} & \textbf{0.079} & \textbf{15} \\ \hline
\multicolumn{1}{l|}{} & \multicolumn{1}{l|}{0.1} & Naive & 0.001 & 0.002 & \textbf{0.052} & \textbf{0.067} & 47 \\
\multicolumn{1}{l|}{} & \multicolumn{1}{l|}{} & Ours & 0.001 & 0.002 & \textbf{0.052} & \textbf{0.067} & \textbf{69} \\ \cline{2-8} 
\multicolumn{1}{l|}{Blog} & \multicolumn{1}{l|}{0.5} & Naive & 0.001 & 0.002 & \textbf{0.052} & \textbf{0.068} & 3 \\
\multicolumn{1}{l|}{} & \multicolumn{1}{l|}{} & Ours & 0.001 & 0.002 & 0.053 & \textbf{0.068} & \textbf{6} \\ \cline{2-8} 
\multicolumn{1}{l|}{} & \multicolumn{1}{l|}{1} & Naive & 0.001 & 0.002 & 0.053 & \textbf{0.068} & 2 \\
\multicolumn{1}{l|}{} & \multicolumn{1}{l|}{} & Ours & 0.001 & 0.001 & \textbf{0.052} & \textbf{0.068} & \textbf{3} 
\end{tabular}
}
    \caption{\textsc{1-Diffractor}}
  \end{subtable}
  \begin{subtable}{\linewidth}
    \centering
    \resizebox{\linewidth}{!}{
    \begin{tabular}{lll|c|c|c|c|c}
 &  &  & $MTI_{seq}T1\downarrow$  & $MTI_{seq}T3\downarrow$  & $MTI_{bow}T1\downarrow$ & $MTI_{bow}T1\downarrow$ & NN$\uparrow$ \\ \hline
\multicolumn{1}{l|}{} & \multicolumn{1}{l|}{500} & Naive & 0.002 & 0.005 & 0.104 & 0.137 & 816 \\
\multicolumn{1}{l|}{} & \multicolumn{1}{l|}{} & Ours & 0.001 & 0.004 & \textbf{0.042} & \textbf{0.135} & \textbf{964} \\ \cline{2-8} 
\multicolumn{1}{l|}{Yelp} & \multicolumn{1}{l|}{1000} & Naive & 0.002 & 0.004 & 0.109 & 0.135 & 342 \\
\multicolumn{1}{l|}{} & \multicolumn{1}{l|}{} & Ours & 0.001 & 0.004 & \textbf{0.043} & \textbf{0.079} & \textbf{936} \\ \cline{2-8} 
\multicolumn{1}{l|}{} & \multicolumn{1}{l|}{1500} & Naive & 0.002 & 0.005 & 0.112 & 0.139 & 203 \\
\multicolumn{1}{l|}{} & \multicolumn{1}{l|}{} & Ours & 0.001 & 0.003 & \textbf{0.051} & \textbf{0.087} & \textbf{879} \\ \hline
\multicolumn{1}{l|}{} & \multicolumn{1}{l|}{500} & Naive & 0.002 & 0.004 & 0.052 & 0.068 & 780 \\
\multicolumn{1}{l|}{} & \multicolumn{1}{l|}{} & Ours & 0.001 & 0.005 & \textbf{0.019} & \textbf{0.035} & \textbf{956} \\ \cline{2-8} 
\multicolumn{1}{l|}{Trustpilot} & \multicolumn{1}{l|}{1000} & Naive & 0.001 & 0.003 & 0.058 & 0.074 & 257 \\
\multicolumn{1}{l|}{} & \multicolumn{1}{l|}{} & Ours & 0.002 & 0.004 & \textbf{0.025} & \textbf{0.041} & \textbf{787} \\ \cline{2-8} 
\multicolumn{1}{l|}{} & \multicolumn{1}{l|}{1500} & Naive & 0.001 & 0.002 & 0.061 & 0.076 & 146 \\
\multicolumn{1}{l|}{} & \multicolumn{1}{l|}{} & Ours & 0.001 & 0.002 & \textbf{0.032} & \textbf{0.049} & \textbf{604} \\ \hline
\multicolumn{1}{l|}{} & \multicolumn{1}{l|}{500} & Naive & 0.001 & 0.002 & 0.042 & 0.058 & 772 \\
\multicolumn{1}{l|}{} & \multicolumn{1}{l|}{} & Ours & 0.001 & 0.002 & \textbf{0.024} & \textbf{0.036} & \textbf{833} \\ \cline{2-8} 
\multicolumn{1}{l|}{Blog} & \multicolumn{1}{l|}{1000} & Naive & 0.001 & 0.001 & 0.044 & 0.056 & 520 \\
\multicolumn{1}{l|}{} & \multicolumn{1}{l|}{} & Ours & 0.001 & 0.001 & \textbf{0.026} & \textbf{0.041} & \textbf{559} \\ \cline{2-8} 
\multicolumn{1}{l|}{} & \multicolumn{1}{l|}{1500} & Naive & 0.001 & 0.002 & 0.049 & 0.061 & \textbf{472} \\
\multicolumn{1}{l|}{} & \multicolumn{1}{l|}{} & Ours & 0.001 & 0.001 & \textbf{0.029} & \textbf{0.045} & 406
\end{tabular}
}
    \caption{\textsc{DP-BART}}
  \end{subtable}
  \label{tab:mia_results}
\end{table}

\subsubsection{Membership Inference Evaluation}
\label{sec:mi}
In the context of privacy-preserving Machine Learning, \textit{Membership Inference Attacks} (MIAs) attempt to infer whether a specific data record (e.g., individual) was part of the data used to train a model \cite{10.1145/3523273}. With textual data, MIAs take on a slightly different meaning, and essentially, the goal of the attacker becomes to infer whether certain textual information was present in the training data \cite{9152761}. To evaluate the resilience of DP text rewriting against MIAs, we run two types of experiments.

\paragraph{Masked Token Inference Attack (MTI)}
Following \citet{chen-etal-2023-customized}, we run a \textit{masked token inference attack}. We leverage the ability of masked language models (MLMs) to predict a masked (hidden) word given the context surrounding the word. Thus, given a privatized text, we test an MLM's ability to predict tokens from the original text when provided with the privatized context. To measure the performance of this attack, we follow a procedure as such:
\begin{enumerate}
    \item For each privatized document, mask each token one by one. In this work, we use \textsc{bert-base-uncased} \cite{devlin-etal-2019-bert}.
    \item Capture the top predictions of the MLM (top-1 and top-3).
    \item Check if the predictions match the exact original token in sequence ($MTI_{seq}$), or if the predictions match any token in the original text ($MTI_{bow}$), as in a bag-of-words.
\end{enumerate}

Thus, for each dataset, we report four scores: $MTI_{seq}T1$, $MTI_{seq}T3$, $MTI_{bow}T1$, and $MTI_{bow}T3$, where T1 and T3 represent considering the top-1 and top-3 predictions, respectively. For all scores, a lower score means higher privacy protection.

\paragraph{Nearest Neighbor Attack (NN)}
We also design a new attack, called the \textit{nearest neighbor attack}, which measures how close (semantically), on average, the privatized text is to the original text given the entire privatized dataset. The procedure is as follows:
\begin{enumerate}
    \item For each document in the original dataset, select this document as the \textit{query} document.
    \item Note the index of the query's private counterpart in the privatized dataset, or the \textit{corpus}.
    \item Using an embedding model and cosine similarity measures, measure for which $k$ value the private document is the $k$-th nearest neighbor to the query document. 
\end{enumerate}
With this, we measure the \textit{plausible deniability} that is created, i.e., the \say{distance} from the original document to the private document. We report the average $k$ over all documents in the private dataset. In the context of MIA, a higher average $k$ would imply higher privacy.

The results of both the MTI and NN experiments on all three privacy datasets can be found in Table \ref{tab:mia_results}.


\begin{table*}[htbp]
  \centering
    \caption{Utility Experiment Results. Utility scores are represented by F1 scores achieved on the corresponding tasks. For all \textsc{1-Diffractor} tasks, we report the per-word $\varepsilon$, as well as the total allocated budget (indicated in parentheses), which is calculated by (\textit{per-word $\varepsilon$})$\cdot$(\textit{Avg. Tokens}), or average number of tokens in a document per dataset. Baseline scores on the non-privatized data are also provided. Note that for \textsc{DP-BART} and the \textit{IMDb} dataset, we take a 20\% random split due to the size of the dataset.}
  \begin{subtable}{\textwidth}
    \centering
    \resizebox{\textwidth}{!}{
    \begin{tabular}{l|cc|cc|cc|cc|cc|cc|cc|cc|cc}
 & \multicolumn{6}{c|}{\textbf{SST2}} & \multicolumn{6}{c|}{\textbf{MRPC}} & \multicolumn{6}{c}{\textbf{MNLI}} \\ \hline
Baseline & \multicolumn{6}{c|}{$96.12_{0.1}$} & \multicolumn{6}{c|}{$86.56_{0.9}$} & \multicolumn{6}{c}{$86.68_{0.2}$} \\ \hline
Avg. Tokens & \multicolumn{6}{c|}{8.31} & \multicolumn{6}{c|}{18.29} & \multicolumn{6}{c}{19.54} \\ \hline
$\varepsilon$ & \multicolumn{2}{c|}{0.1 (0.83)} & \multicolumn{2}{c|}{0.5 (4.16)} & \multicolumn{2}{c|}{1.0 (8.31)} & \multicolumn{2}{c|}{0.1 (1.83)} & \multicolumn{2}{c|}{0.5 (9.15)} & \multicolumn{2}{c|}{1.0 (18.29)} & \multicolumn{2}{c|}{0.1 (1.95)} & \multicolumn{2}{c|}{0.5 (9.77)} & \multicolumn{2}{c}{1.0 (19.54)} \\ \hline
Distribution & Naive & Ours & Naive & Ours & Naive & Ours & Naive & Ours & Naive & Ours & Naive & Ours & Naive & Ours & Naive & Ours & Naive & Ours \\ \hline
Utility (F1) $\uparrow$ & $80.40_{0.3}$ & $78.93_{0.2}$ & $87.21_{0.1}$ & $85.31_{0.2}$ & $91.20_{0.1}$ & $88.94_{0.1}$ & $70.12_{1.4}$ &  $64.67_{0.1}$ &  $75.93_{1.1}$ & $74.21_{1.8}$ & $71.39_{2.3}$ & $73.48_{6.6}$ & $40.37_{6.6}$ & $39.78_{5.7}$ & $56.74_{12.8}$ & $55.30_{4.5}$ & $67.77_{0.9}$ & $67.26_{2.0}$ \\
CS $\uparrow$ & 0.508 & 0.459 & 0.707 & 0.625 & 0.813 & 0.721 & 0.438 & 0.382 & 0.665 & 0.573 & 0.786 & 0.689 & 0.464 & 0.406 & 0.706 & 0.610 & 0.819 & 0.719 \\
BLEU $\uparrow$ & 0.163 & 0.155 & 0.240 & 0.203 & 0.341 & 0.267 & 0.126 & 0.118 & 0.205 & 0.174 & 0.314 & 0.243 & 0.164 & 0.157 & 0.261 & 0.220 & 0.372 & 0.297 \\
PPL $\downarrow$ & 9372 & 9849 & 6329 & 7266 & 5152 & 5813 & 1808 & 1912 & 991 & 1233 & 552 & 795 & 1892 & 2035 & 1361 & 1232 & 648 & 857
\end{tabular}
}
    \caption{1-Diffractor (1/2)}
  \end{subtable}
  \begin{subtable}{\textwidth}
    \centering
    \resizebox{\textwidth}{!}{
    \begin{tabular}{l|cc|cc|cc|cc|cc|cc|cc|cc|cc}
 & \multicolumn{6}{c|}{\textbf{BBC}} & \multicolumn{6}{c|}{\textbf{DocNLI}} & \multicolumn{6}{c}{\textbf{IMDb}} \\ \hline
Baseline & \multicolumn{6}{c|}{$98.73_{0.3}$} & \multicolumn{6}{c|}{$87.82_{0.4}$} & \multicolumn{6}{c}{$95.84_{0.2}$} \\ \hline
Avg. Tokens & \multicolumn{6}{c|}{399.34} & \multicolumn{6}{c|}{285.22} & \multicolumn{6}{c}{225.97} \\ \hline
$\varepsilon$ & \multicolumn{2}{c|}{0.1 (39.99)} & \multicolumn{2}{c|}{0.5 (199.97)} & \multicolumn{2}{c|}{1.0 (399.94)} & \multicolumn{2}{c|}{0.1 (28.52)} & \multicolumn{2}{c|}{0.5 (142.61)} & \multicolumn{2}{c|}{1.0 (285.22)} & \multicolumn{2}{c|}{0.1 (22.60)} & \multicolumn{2}{c|}{0.5 (112.99)} & \multicolumn{2}{c}{1.0 (225.97)} \\ \hline
Distribution & Naive & Ours & Naive & Ours & Naive & Ours & Naive & Ours & Naive & Ours & Naive & Ours & Naive & Ours & Naive & Ours & Naive & Ours \\ \hline
Utility (F1) $\uparrow$ & $92.17_{0.6}$ &  $75.87_{15.6}$ & $95.87_{0.0}$ & $95.56_{0.3}$ & $96.83_{0.4}$ & $95.77_{1.6}$ & $79.10_{0.5}$ & $78.30_{0.5}$  & $82.02_{0.2}$ & $80.34_{1.0}$ & $82.86_{0.3}$ & $82.02_{0.8}$ & $57.31_{9.9}$ & $77.64_{14.8}$ & $92.73_{0.2}$ & $92.77_{0.1}$ & $94.46_{0.0}$ & $93.94_{0.1}$ \\
CS $\uparrow$ & 0.507 & 0.461 & 0.753 & 0.699 & 0.855 & 0.807 & 0.593 & 0.546 & 0.814 & 0.765 & 0.879 & 0.845 & 0.471 & 0.421 & 0.707 & 0.639 & 0.818 & 0.750 \\
BLEU $\uparrow$ & 0.132 & 0.127 & 0.222 & 0.196 & 0.330 & 0.279 & 0.112 & 0.107 & 0.190 & 0.167 & 0.248 & 0.218 & 0.167 & 0.162 & 0.263 & 0.234 & 0.370 & 0.317 \\
PPL $\downarrow$ & 692 & 725 & 324 & 397 & 173 & 236 & 716 & 795 & 311 & 376 & 212 & 257 & 621 & 655 & 309 & 371 & 186 & 240
\end{tabular}
}
    \caption{1-Diffractor (2/2)}
  \end{subtable}
  \begin{subtable}{\textwidth}
    \centering
    \resizebox{\textwidth}{!}{
    \begin{tabular}{l|cc|cc|cc|cc|cc|cc|cc|cc|cc}
 & \multicolumn{6}{c|}{\textbf{BBC}} & \multicolumn{6}{c|}{\textbf{DocNLI}} & \multicolumn{6}{c}{\textbf{IMDb*}} \\ \hline
Baseline & \multicolumn{6}{c|}{$98.73_{0.3}$} & \multicolumn{6}{c|}{$87.82_{0.4}$} & \multicolumn{6}{c}{$95.13_{0.2}$} \\ \hline
Avg. Tokens & \multicolumn{6}{c|}{399.34} & \multicolumn{6}{c|}{285.22} & \multicolumn{6}{c}{225.97} \\ \hline
$\varepsilon$ & \multicolumn{2}{c|}{500} & \multicolumn{2}{c|}{1000} & \multicolumn{2}{c|}{1500} & \multicolumn{2}{c|}{500} & \multicolumn{2}{c|}{1000} & \multicolumn{2}{c|}{1500} & \multicolumn{2}{c|}{500} & \multicolumn{2}{c|}{1000} & \multicolumn{2}{c}{1500} \\ \hline
Distribution & Naive & Ours & Naive & Ours & Naive & Ours & Naive & Ours & Naive & Ours & Naive & Ours & Naive & Ours & Naive & Ours & Naive & Ours \\ \hline
Utility (F1) $\uparrow$ & $40.53_{1.8}$ & $32.17_{1.2}$ & $90.48_{0.4}$ & $33.86_{1.3}$ & $92.59_{0.1}$ & $32.59_{0.8}$ & $66.96_{2.3}$ & $75.89_{0.8}$ & $75.71_{0.9}$ &  $74.36_{2.1}$ & $75.46_{0.9}$ & $76.29_{0.8}$ & $70.23_{1.3}$ &  $49.23_{1.0}$ & $87.77_{0.6}$ &  $50.20_{1.9}$ & $89.07_{0.3}$ & $53.37_{2.4}$ \\
CS $\uparrow$ & 0.145 & 0.065 & 0.414 & 0.072 & 0.489 & 0.072 & 0.168 & 0.069 & 0.453 & 0.104 & 0.531 & 0.176 & 0.187 & 0.091 & 0.411 & 0.102 & 0.454 & 0.129 \\
BLEU $\uparrow$ & 0.017 & 0.000 & 0.051 & 0.000 & 0.065 & 0.000 & 0.012 & 0.000 & 0.038 & 0.001 & 0.047 & 0.004 &  0.028 & 0.000 & 0.075 & 0.002 & 0.090 & 0.006 \\
PPL $\downarrow$ & 23 & 34667 & 11 & 15207 & 11 & 5811 & 27 & 18314 & 12 & 8303 & 11 & 4936 & 19 & 25054 & 9 & 5351 & 9 & 1908
\end{tabular}
}
    \caption{DP-BART}
  \end{subtable}
  \label{tab:utility}
\end{table*}

\subsection{Utility Experiments}
In addition to measuring the privacy-preserving capabilities of the DP rewriting methods both with and without budget distribution, we also measure the utility preservation, namely the effect on downstream task utility between a naive and our proposed distribution.

\subsubsection{Datasets and Tasks}
We use six datasets for utility evaluation.

\paragraph{\textsc{GLUE} Datasets}
The \textsc{GLUE} Benchmark \cite{wang-etal-2018-glue} consists of nine datasets focused on evaluating the general language understanding capabilities of language models. We choose three datasets from each of the three sub-tasks of the benchmark: \textsc{SST-2} (sentiment analysis), \textsc{MRPC} (sentence similarity), and \textsc{MNLI} (textual entailment). In the case of the MNLI, we take a 10\% subset for a total of 39,270 training instances. Note that since these datasets are all a maximum of one sentence, we only evaluate them with \textsc{1-Diffractor}.

\paragraph{BBC News}
The BBC News dataset\footnote{\url{http://mlg.ucd.ie/datasets/bbc.html}} is a collection of 3147 news articles from the BBC platform, where each news article belongs to one of five popular news categories: \textit{business}, \textit{entertainment}, \textit{politics}, \textit{sports}, or \textit{tech}. This creates a five-class classification task.

\paragraph{DocNLI}
The \textit{DocNLI} dataset \cite{yin-etal-2021-docnli} introduces a \textit{document}-level entailment prediction task. The dataset consists of (\textit{premise}, \textit{hypothesis}) pairs, each marked as \textit{entailment} or \textit{not entailment}. As the original dataset is very large, we take a 1\% random sample, resulting in a 9136-row dataset with two classes. 

\paragraph{IMDb Reviews}
The IMDb Dataset \cite{maas-etal-2011-learning} consists of 50k movie reviews from the IMDb platform, each labeled as positive or negative. 

\subsubsection{Utility Evaluation}
To evaluate utility, we follow a similar procedure to the model training described in Section \ref{sec:ep_eval}. Firstly, each dataset is privatized using our two chosen DP rewriting methods, their respective three $\varepsilon$ values, and the two distribution techniques. The resulting datasets, including the original baseline datasets, are used to train a \textsc{deberta-v3-base} classification model for one epoch.

For each training procedure, we report the F1 score achieved by the trained model on the 10\% held-out test set. These results are found in Table \ref{tab:utility}. In addition, we report three other metrics to capture the quality and coherence of the privatized text documents:
\begin{enumerate}
    \item \textit{Cosine Similarity} (CS): we measure the average cosine similarity of the embeddings of the original texts and their privatized counterparts, using Sentence Transformers \cite{reimers-gurevych-2019-sentence}, specifically with the \textsc{all-MiniLM-L12-v2} embedding model.
    \item \textit{BLEU}: the bilingual evaluation understudy (BLEU) score is used to measure the quality of generated text (i.e., private texts) as compared to a reference text (i.e., original texts). We report the average BLEU score using the \textsc{nltk} package.
    \item \textit{Perplexity} (PPL): \textit{perplexity} can be used as a proxy to measure the coherence and understandability of a text, as it measures how \say{surprised} a language model is when seeing a given text. We use a \textsc{GPT-2} model to measure perplexity \cite{10.1145/3485447.3512232, meisenbacher-matthes-2024-thinking}. For performance, we limit input texts to the first 256 tokens.
\end{enumerate}


\begin{table*}[htbp]
  \centering
    \caption{Ablation Study Results. Baseline scores represent the results using all five distribution methods, while \textit{/X} denotes the usage of four methods without \textit{X}. Non-baseline values indicate the relative change ($\Delta$, in \%) from the baseline. Note that ablation results from the \textit{masked token inference} (MTI) evaluation are not reported, due to non-significant changes. *$std = 13.1$}
  \begin{subtable}{\textwidth}
    \centering
    \resizebox{\textwidth}{!}{
    \begin{tabular}{l|cccccccccccc|cccccc}
\multicolumn{1}{r|}{$\epsilon$} & \multicolumn{6}{c|}{0.1} & \multicolumn{6}{c|}{0.5} & \multicolumn{6}{c}{1.0} \\ \cline{2-19} 
 & \multicolumn{1}{c}{Baseline} & /IC & /POS & /NER & /WI & \multicolumn{1}{c|}{/SD} & Baseline & /IC & /POS & /NER & /WI & SD & Baseline & /IC & /POS & /NER & /WI & /SD \\ \hline
Utility (F1) $\uparrow$ ($\Delta$) & $93.53_{0.0}$ & +0.00 & +0.00 & +0.00 & +0.00 & \multicolumn{1}{c|}{+0.17} & $94.01_{0.9}$ & +0.21 & -0.18 & -0.48 & -0.48 & -0.48 & $94.45_{0.6}$ & -0.81 & -0.92 & -0.46 & -0.92 & -0.92 \\
Adv. F1 (s) $\downarrow$ ($\Delta$)& 42.37 & -1.56 & -1.68 & +0.34 & -3.12 & \multicolumn{1}{c|}{-3.55} & 55.09 & -0.29 & -0.52 & +1.96 & -0.99 & -2.14 & 62.60 & -0.87 & +0.52 & +1.56 & -2.25 & -2.08 \\
Adv. F1 (a) $\downarrow$ ($\Delta$) & $82.35_{0.9}$ & +0.56 & +1.73 & +1.79 & +3.97 & \multicolumn{1}{c|}{+0.04} & $88.44_{1.4}$ & +1.68 & +2.52 & +3.41 & -2.56 & +0.42 & $92.87_{0.7}$ & -0.60 & -0.94 & -0.58 & -1.16 & -2.85 \\
$\gamma$ (s) $\uparrow$ ($\Delta$)& 1.81 & +0.02 & +0.03 & -0.00 & +0.05 & \multicolumn{1}{c|}{-1.02} & 1.31 & -0.12 & +0.12 & -0.02 & +0.32 & +0.34 & 0.91 & +0.53 & +0.59 & +0.28 & +0.63 & +0.63 \\
$\gamma$ (a) $\uparrow$ ($\Delta$) & 1.21 & -0.01 & -0.03 & -0.03 & -0.06 &  \multicolumn{1}{c|}{-0.11} & 0.81 & -0.17 & +0.07 & +0.04 & +0.34 & +0.30 & 0.46 & +0.52 & +0.60 & +0.02 & +0.60 & +0.63 \\ \hline
$NN$  $\uparrow$ ($\Delta$)& 154 & +4 & -3 & -27 & +29 & \multicolumn{1}{c|}{+32} & 17 & +0 & -2 & -6 & +6 & +6 & 4 & +0 & +0 & -1 & +1 & +3
\end{tabular}
}
    \caption{Yelp}
  \end{subtable}
  \begin{subtable}{\textwidth}
    \centering
    \resizebox{\textwidth}{!}{
       \begin{tabular}{l|lccccccccccc|cccccc}
\multicolumn{1}{r|}{$\epsilon$} & \multicolumn{6}{c|}{0.1} & \multicolumn{6}{c|}{0.5} & \multicolumn{6}{c}{1.0} \\ \cline{2-19} 
 & \multicolumn{1}{c}{Baseline} & /IC & /POS & /NER & /WI & \multicolumn{1}{c|}{/SD} & Baseline & /IC & /POS & /NER & /WI & SD & Baseline & /IC & /POS & /NER & /WI & /SD \\ \hline
Utility (F1) $\uparrow$ ($\Delta$) &  $78.93_{0.2}$ & +0.23 & +0.79 & +0.37 & +0.40 & \multicolumn{1}{c|}{-1.09} & $85.31_{0.2}$ & -9.31* & -1.03 & +0.96 & +0.43 & -1.54 & $88.94_{0.1}$ & +0.05 & +0.67 & +0.82 & +0.48 & -3.28 \\
CS $\uparrow$ ($\Delta$) & 0.459 & +0.001 & +0.010 & +0.011 & +0.001 & \multicolumn{1}{c|}{-0.028} & 0.625 & -0.001 & +0.017 & +0.019 & -0.002 & +0.058 & 0.721 & +0.001 & +0.022 & +0.022 & -0.001 & +0.074 \\
BLEU $\uparrow$ ($\Delta$)& 0.155 & +0.000 & +0.001 & +0.003 & +0.000 & \multicolumn{1}{c|}{-0.005} & 0.203 & +0.000 & +0.007 & +0.010 & -0.002 & -0.019 & 0.267 & +0.001 & +0.015 & +0.023 & -0.004 & -0.40 \\
PPL $\downarrow$ ($\Delta$) & 9849 & +417 & +514 & +658 & -2 & \multicolumn{1}{c|}{+330} & 7266 & +251 & -259 & -194 & +470 & +621 & 5813 & +214 & -142 & +189 & +77 & +943
\end{tabular}
}
    \caption{SST2}
  \end{subtable}
  \label{tab:ablation}
\end{table*}

\subsection{Ablation Study}
\label{sec:ablation}
The final component of our experiments involves an ablation study with our privacy budget distribution toolkit, namely to measure the individual effect of the five proposed scoring methods. Thus, we are able to identify which of the methods leads to higher privacy and utility preservation, and which may need future improvement.

\subsubsection{Setup}
For the ablation study, we focus on one mechanism (\textsc{1-Diffractor}) and two datasets (\textit{SST2} and \textit{Yelp}). For each dataset, we privatize all documents under the same setup as in the previous experiments, i.e., with the base epsilons of $\varepsilon \in \{0.1, 0.5, 1\}$, scaled to the average number of tokens per document. However, as opposed to before, we privatize each (dataset, $\varepsilon$) pair \textit{five} times, each time with one method from our distribution toolkit disabled.

For the utility (\textit{SST2}) and privacy (\textit{Yelp}) ablation, we report the change in score ($\Delta$), or how the corresponding score was affected by the disabling of the particular distribution scoring method. Thus, a more negative change in a metric, e.g., loss in utility, would imply that a given method is more effective when included than disabled.

The results of the ablation study are presented in Table \ref{tab:ablation}.

\section{Discussion}
We critically reflect on the results presented in this work, as well as discuss opportunities and recommendations based on our findings.

\subsection{More Privacy, Same Budget}
An analysis of the experimental results begins with the strengths exhibited when performing DP text rewriting under our proposed distribution scheme rather than an equal, \say{naive} distribution.

As showcased in Table \ref{tab:ep_results}, using our toolkit leads to stronger protection against adversaries in attribute inference attacks (gender or authorship), in 15/18 static attacker scenarios and 13/18 adaptive attacker scenarios. These results are echoed in the Membership Inference evaluations, where our distribution outperforms naive distribution in nearly all of the $MTI$ results, as well as all but one $NN$ score. These results support the hypothesis that a more informed spending of the privacy budget in DP text rewriting can afford higher privacy levels given the same overall budget.

The implications of these results are clear. The provision of a certain privacy budget for a document leads to a particular privacy guarantee on paper (i.e., a DP guarantee), yet the \textit{empirical} effects of such a guarantee can differ significantly depending on how the budget is \say{spent}. These results show that in DP text rewriting, simply choosing an $\varepsilon$ budget is not enough -- a careful consideration of how this budget is allocated must also take place in order to maximize privacy protections in practice.

\subsection{The Privacy-Utility Trade-off in Action}
Naturally, a discussion of the privacy protections that our distribution method bolsters must also be discussed in light of its effect on the utility of the privatized data, or the \textit{privacy-utility trade-off}.

In Table \ref{tab:utility}, a clear decrease in utility can be observed in nearly all cases of our versus naive distribution. On the surface, this utility loss is to be expected: if our aim is to privatize texts more rigorously by focusing on certain component tokens more than others, this will inevitably lead to a weaker semantic signal from the data. Interestingly, we observe that the effect on utility is different for our two chosen mechanisms. In this case of a word-level mechanism (\textsc{1-Diffractor}), the utility loss stays consistent and always entails a rather small loss. In the case of \textsc{DP-BART}, the effect on utility is clearly more severe, as demonstrated in the case of \textit{BBC} and \textit{IMDb}. This significant loss in utility is not absolute, though, as can be showcased in the \textit{DocNLI} experiments, where our distribution with \textsc{DP-BART} performs the same or better than a naive distribution. These results highlight that budget distribution is not as clear-cut with document-level DP mechanisms, where segmenting inputs into sentences yields varying degrees of output quality.

In light of the infamous \textit{privacy-utility trade-off}, we see that the relative consistency of the distributed \textsc{1-Diffractor} utility loss is met with a generally lower capability to mitigate privacy risks. In particular, the results in Tables \ref{tab:ep_results} and \ref{tab:mia_results} show that data rewritten with \textsc{1-Diffractor}, whether distributed with our method or not, is not as strong in protection against attribute or membership inference attacks. On the other hand, while \textsc{DP-BART}, particularly when distributed, can largely neutralize any privacy threat, the effect on utility is so significant that the trade-offs may be more similar to \textsc{1-Diffractor} than meets the eye.

The trade-offs are most clearly demonstrated in Table \ref{tab:ep_results} with the \textit{Relative Gain} ($\gamma$) metric, which tells an interesting story. In defending against authorship attribution (\textit{Yelp}), our distribution method nearly always (11/12) leads to more favorable trade-offs, and regardless of distribution method, \textsc{DP-BART} yields significantly higher relative gains. In contrast, the findings with gender identification (\textit{Trustpilot}) are more mixed, with no clear winner regarding mechanism or distribution method. Beyond showing the complexity of the privacy-utility trade-off, these findings imply that considerations of budget distribution are also \textit{task-specific}, and gains in terms of the trade-off do not come uniformly across different tasks.

\begin{table*}[htbp]
\footnotesize
  \centering
    \caption{Selected Examples of Rewritten Texts from the \textit{Yelp} Dataset, using the \textsc{DP-BART} Mechanism.}
    \resizebox{\textwidth}{!}{
   \begin{tabular}{cc|c|p{0.8\linewidth}}
\multicolumn{1}{r|}{} & $\epsilon$ & \multicolumn{1}{r}{Original:} & My 8 year old just LOVES it here - from musical instruments to jewelry to hand bags, everything is giftable and comes from a fair-trade community.  It's great fun - but a bit pricey - and nothing here is a neccesity - but if you gotta buy gifts, at least here they are unique and helping another community.  We love it! \\ \hline
\multicolumn{1}{c|}{\multirow{6}{*}{\rotatebox[origin=c]{90}{DP-BART}}} & 
\multicolumn{1}{c|}{\multirow{2}{*}{1000}} & Naive & My 8 year old just loves it here - it's a great place to shop for gifts and toys - and the kids love it here! - from the start of the year - we have a lot of fun with it! - and it is a bit pricey - but if you gotta buy something here, it's worth it.My 8 yr old just LOVES it here \\
\multicolumn{1}{c|}{} & \multicolumn{1}{c|}{} & Ours & Just a small-b\%\%\% -\%\% e-the-w\%\%-\%\% The e-all-s\%\% It's also a great place to be if you want to be a part of the community - but it's also very difficult to be in the community.It's a great complete \\ \cline{2-4} 
\multicolumn{1}{c|}{} & \multicolumn{1}{c|}{\multirow{2}{*}{1500}} & Naive & My 8 year old loves this store - it's a gift shop that has everything you need for your kids. The kids are all over the place. The store is small - but the kids love it.My 8 yr old just LOVES this store. It's a great gift shop. The prices are great - but if you gotta buy a gift, this is a great \\
\multicolumn{1}{c|}{} & \multicolumn{1}{c|}{} & Ours & This is the way we are going to go this year.We have a little bit of a way of doing this. The way we do it It's a bit pricey - but if you can't afford it, it's not a bad thing - and nothing here is cheap - but it's a little bit of a wonders
\end{tabular}
}
  \label{tab:examples_small}
\end{table*}

\subsection{When Does Distribution Make Sense? A Qualitative Analysis}
Beyond the reported metrics in this work, one can look at side-by-side examples of DP-rewritten texts for insights.

Looking at selected examples in Table \ref{tab:examples_small}, one can begin to observe the differences in rewritten texts between the two distribution schemes. In the \textsc{DP-BART} example with $\varepsilon = 1500$, the naive distribution method not only fails to hide the \say{8 year old} cue, but it also magnifies this phrase in a later sentence. On the other hand, the rewritten text with our method makes no mention of this phrase, and in the case of $\varepsilon=1000$, our rewritten text completely \say{masks} out the original sentence, albeit with a non-coherent replacement. Similarly, in the case of \textsc{1-Diffractor}, we notice that more words are perturbed (changed) in the distributed examples rather than the naive. Moreover, certain writing cues, such as \say{We love it!}, are never privatized in the naive distribution, but are finally considered in our distribution scheme (at $\varepsilon = 0.1$).

While these insights are anecdotal evidence, we hold that such differences are important in the consideration of text privatization. The examples illustrate the fact that in any given text, not all components of the greater whole are equally important, both semantically and from a privacy point of view; therefore, privacy budget allocation should follow the same logic. At the same time, the examples also demonstrate the pitfalls of a more informed distribution, such as in the non-coherent outputs of \textsc{DP-BART} at lower budgets. In addition, even with our proposed distribution, mechanisms such as \textsc{1-Diffractor} struggle with truly obfuscating the original text, as in any case, significant semantic cues still remain. The qualitative analysis, therefore, teaches that while there is sense in distributing the privacy budget intelligently, there is still much work to be done. For more examples, we refer the reader to Table \ref{tab:examples}.

\subsection{Investigating the Distribution Methods}
Following our ablation study (Table \ref{tab:ablation}), we critically reflect on the merits and limitations of the individual distribution methods in our toolkit, leading to ideas and suggestions for further improvement.

\paragraph{Methods that impact utility}
We observe that, with slight deviations, all methods besides \textbf{SD} lead to an increase in utility when removed from the evaluations, thus suggesting that the utilization of these distribution methods leads to lower utility of the data. In the interesting case of \textbf{SD}, disabling this method actually leads to quite significant drops in performance (see the utility scores of \textit{SST2} in the Table \ref{tab:ablation}), which can plausibly be attributed to this method removing \say{outlier} tokens in the texts that may overtrain models to particular tokens. The utility-boosting properties of this method are also demonstrated in the other utility metrics, where, for example, disabling \textbf{SD} leads to the largest decreases in CS and BLEU.

\paragraph{Methods that impact privacy}
The privacy results of the ablation study uncover an interesting dichotomy. In general, disabling any method besides \textbf{NER} seems to \textit{improve} (lower) privacy scores in the \textit{static} (s) attacker setting, whereas disabling \textbf{NER} always leads to \textit{worse} (higher) results. This suggests that focusing the privacy budget on named entities is important in the static attacker setting.

In the adaptive attacker setting, other interesting findings arise. At lower privacy budgets (i.e., $\varepsilon = 0.1$ and $\varepsilon = 0.5$), \textit{all} methods play an important role in reducing adversarial performance, except for one case (\textbf{WI} at $\varepsilon = 0.5$). However, at the higher budget setting ($\varepsilon = 1$), removing any given method only serves to improve (lower) the privacy results. This implies that budget distribution is most important in lower privacy budget regimes, where defending against more capable adversaries necessitates careful allocation of $\varepsilon$.

Similarly, the effect of certain methods is pronounced with lower privacy budgets, as showcased by the \textit{NN} ablation scores. Here, we observe that \textbf{WI} and \textbf{SD} are influential against membership inference, whereas others do not play as large of a role.

As a final note regarding the low versus high privacy budgets, the relative gains ($\gamma$) of Table \ref{tab:ablation} illustrate that as the overall budget increases, it may make less sense (from a trade-off perspective) to distribute the budget with our method. This, however, would largely depend on whether balancing the trade-off is more important, as opposed to optimizing privacy (e.g., membership inference).

\paragraph{Main Takeaways}
The results of the ablation study, in conjunction with the other results we present, show promise in the optimization of privacy budgets in DP text rewriting, while also highlighting important considerations going forward.

Our experiments present a cursory overview of the potential effectiveness of our proposed budget distribution methods, but the results merit further investigations. Taking \textbf{POS} as an example, we observe in Table \ref{tab:ablation} that this method generally contributes to better privacy scores, as showcased by the loss of privacy when it is disabled. However, this is met with increases in utility in some settings, and decreases in others. In this particular example, we cannot say with certainty whether the fixed weighting scheme of \textbf{POS} is optimal, and furthermore, exactly which weights can be adjusted. This discussion leads to the further consideration that while it is plausible that certain parts of speech are more relevant to privatization than others, we simply do not have the data to produce a more intelligent weighting scheme. This observation extends to our other proposed methods in the toolkit, where our initial assumptions about what is important in text privatization would be well-served to be backed by more informed data.

The various results presented in this work give credence to the complexity of privacy in textual data, as the many dimensions we present (i.e., the multiple angles of privacy and utility) make it difficult to definitively judge effectiveness in privatization. While this naturally calls for more work in privacy benchmarking and privacy metrics, it also sheds light on the subjective and individual nature of privacy in text. As an example, if privacy is strictly important in a certain data sharing scenario and one wishes to protect against strong adversaries, our budget distribution methods would be a very sensible choice. On the other hand, if utility is crucial while privacy is secondary, using a higher privacy budget without distribution might be the wiser choice. Although a continuation of this discussion is outside the scope of this work, the empirical results shown here certainly beckon for further such conversations.

\section{Related Work}
The field of DP text rewriting can be traced back to earlier works on authorship obfuscation using word-level Metric Differential Privacy \cite{fernandes2019generalised}. Other works focusing on DP in NLP sought to improve word-level mechanisms, with later works tackling the challenges of utility preservation or efficiency \cite{weggenmann2018syntf, xu2021density, yue, Chen2022ACT, carvalho2023tem, 10.1145/3643651.3659896}. Later works transitioned to higher levels of syntactic hierarchy, such as with sentences \cite{Meehan2022SentencelevelPF} or document-level latent representations of text \cite{Bo2019ERAEDP, 10.1145/3485447.3512232, igamberdiev-habernal-2023-dp, igamberdiev-etal-2024-dp}. DP text rewriting methods leveraging generative language models \cite{mattern-etal-2022-limits, utpala-etal-2023-locally, flemings-annavaram-2024-differentially} have also been proposed in the recent literature as a way to produce more coherent privatized texts. 

Researchers have also focused on identifying and addressing challenges in the field, especially at the core, where the integration of DP into the NLP realm is not immediately straightforward \cite{habernal-2021-differential,klymenko-etal-2022-differential}. Beyond clear challenges in the generation of coherent and utility-preserving privatized text \cite{feyisetan2021research,mattern-etal-2022-limits}, questions of benchmarking and reproducibility \cite{igamberdiev-etal-2022-dp, meisenbacher-etal-2024-comparative} have also been raised as important paths for future research. Finally, the meaning behind the guarantees that DP rewriting provides has also been a point of investigation \cite{vu-etal-2024-granularity,meisenbacher-matthes-2024-thinking}.

\section{Conclusion}
In this work, we investigate methods to improve the effectiveness of DP text rewriting by focusing on a more informed distribution of privacy budget amongst the tokens of a document. Given an input document and a fixed $\varepsilon$ budget, we propose five methods and a scoring scheme to determine a sensible allocation of the budget to each of the document's components. In our conducted privacy experiments, we learn that in many cases, our proposed budget distribution leads to higher preserved privacy, against both attribute and membership inference attacks. At the same time, we observe that enhanced privacy does not come for free, as our budget distribution largely leads to lower utility in the privatized data.

Our findings highlight the importance of a more intelligent consideration of how a privacy budget is spent in DP text rewriting, resting upon the hypothesis that not all aspects of a text are equally as privacy-sensitive. We empirically demonstrate the privacy-utility trade-off at work, as well as qualitatively analyze the effects of budget distribution. Above all, our findings reveal that much work remains towards designing an optimal budget allocation scheme, and our proposed methods provide the groundwork for doing so.

As such, we propose that future work continues the discussion on the merits and challenges of informed privacy budget distribution in DP text rewriting. In particular, we hope that our proposed methods can be fine-tuned for better privacy protection, which would ideally be supported by user studies and a greater understanding of what it means to preserve privacy in textual data. Additionally, the extension of our work, both in distribution methods and rigorous testing on more DP mechanisms, would help to broaden the initial findings we present in this work.


\bibliographystyle{ACM-Reference-Format}
\balance
\bibliography{sample-base}

\newpage
\appendix
\section{Pseudocode}
Algorithms \ref{alg:ic}, \ref{alg:word_importance}, and \ref{alg:sentence_diff} contain pseudocode for three of the five introduced scoring methods for privacy budget distribution.

\begin{algorithm}[H]
\caption{Information Content Scoring}
    \label{alg:ic}
    \begin{algorithmic}
        \Require tokens $t = t_1...t_n$, \\IC corpora $IC = IC_1...IC_n$, \\WordNet $wn$
        \Ensure IC scores $s = s_1...s_n$

        \State $\texttt{s} \gets []$
        \For {$idx \in 1...len(t)$}
            \State $\texttt{tt} \gets \textit{lemmatize}(t_{idx})$
            \State $\texttt{synsets} \gets wn(\texttt{tt})$
            \State $\texttt{scores} \gets []$
            \For {$\texttt{ic} \in IC$}
                \If {all(\texttt{synsets} not in ic)}
                    \State $\texttt{scores}.\textit{append}(1.0)$
                    \State $\textit{continue}$
                \EndIf
            
                \State $\texttt{score} = \textit{get\_max\_score}(\textit{ic}, \texttt{synsets})$
                \State $\texttt{scores}.\textit{append}(\texttt{score})$
            \EndFor
            \State $\texttt{s}.\textit{append}(\textit{mean}(\texttt{scores}))$
        \EndFor
        \State \Return $s$
    \end{algorithmic}
\end{algorithm}

\begin{algorithm}[H]
\caption{Word Importance Scoring}
    \label{alg:word_importance}
    \begin{algorithmic}
        \Require text $\mathcal{D}$ \\ tokens $t = t_1...t_n$, \\ embedding model $\mathcal{M}$
        \Ensure Importance scores $s = s_1...s_n$

        \State $\texttt{s} \gets []$
        \For {$idx \in 1...len(t)$}
            \State $\texttt{remainder} \gets \mathcal{D} \setminus t_{idx}$
            \State $\texttt{sim} \gets cosine\_similarity(\mathcal{M}, t_{idx}, \texttt{remainder})$
            \State $\texttt{score} \gets 1 - \texttt{sim}$
            \State $\texttt{s}.\textit{append}(\texttt{score})$
        \EndFor
        \State \Return $s$
    \end{algorithmic}
\end{algorithm}

\begin{algorithm}[H]
\caption{Sentence Difference Scoring}
    \label{alg:sentence_diff}
    \begin{algorithmic}
        \Require text $\mathcal{D}$ \\ tokens $t = t_1...t_n$, \\ embedding model $\mathcal{M}$
        \Ensure Importance scores $s = s_1...s_n$

        \State $\texttt{s} \gets []$
        \For {$idx \in 1...len(t)$}
            \State $\texttt{remainder} \gets \mathcal{D} \setminus t_{idx}$
            \State $\texttt{sim} \gets cosine\_similarity(\mathcal{M}, \mathcal{D}, \texttt{remainder})$
            \State $\texttt{score} \gets \texttt{sim}$
            \State $\texttt{s}.\textit{append}(\texttt{score})$
        \EndFor
        \State \Return $s$
    \end{algorithmic}
\end{algorithm}

\section{Examples}
Table \ref{tab:examples} provides a full set of examples from one text in the \textit{Yelp} dataset. Most importantly, texts privatized with both naive distribution and our proposed method are shown.

\begin{table*}[htbp]
\small
  \centering
    \caption{Examples of Rewritten Texts from the \textit{Yelp} Dataset.}
    \resizebox{\textwidth}{!}{
   \begin{tabular}{cc|c|p{0.8\linewidth}}
\multicolumn{1}{r|}{} & $\epsilon$ & \multicolumn{1}{r}{Original:} & My 8 year old just LOVES it here - from musical instruments to jewelry to hand bags, everything is giftable and comes from a fair-trade community.  It's great fun - but a bit pricey - and nothing here is a neccesity - but if you gotta buy gifts, at least here they are unique and helping another community.  We love it! \\ \hline
\multicolumn{1}{c|}{\multirow{6}{*}{\rotatebox[origin=c]{90}{1-Diffractor}}} & \multicolumn{1}{c|}{\multirow{2}{*}{0.1}} & Naive & my 5 today condition just judged it here - from soprano accordion to jewellery to bending bags, anybody is giftable and comes from a fair-trade science . it's good enjoy - but a bit darned - and that here is a neccesity - but if you get rbis deals gifts, at explores here they are unique and warmth another organizations . we love it! \\
\multicolumn{1}{c|}{} & \multicolumn{1}{c|}{} & Ours & my 2 begun grader just like it here - from cavern hula to reclaiming to leg bags, everything is giftable and block from a fair-trade dabbling . it's witnessed crafts - but a splinter overpriced - and nothing here is a neccesity - but if you always ai purchased naive, at block here they are compare and aid another marten . we enjoyed it! \\ \cline{2-4} 
\multicolumn{1}{c|}{} & \multicolumn{1}{c|}{\multirow{2}{*}{0.5}} & Naive & my 6 year old just adores it here - from musical strings to jewelry to arm bags, things is giftable and comes from a fair-trade community . it's great fun - but a bit pricey - and nada here is a neccesity - but if you gotta buy gifts, at about here they are unique and helping another community . we love it! \\
\multicolumn{1}{c|}{} & \multicolumn{1}{c|}{} & Ours & my 7 july replaced just loves it here - from musical playback to jewellery to hold suitcases, everything is giftable and follows from a fair-trade community . it's good fun - but a bit pricey - and nothing here is a neccesity - but if you gotten ai buy baskets, at least here they are unique and helping another societies . we love it! \\ \cline{2-4} 
\multicolumn{1}{c|}{} & \multicolumn{1}{c|}{\multirow{2}{*}{1}} & Naive & my 8 year old just loves it here - from musical instruments to antique to hand bags, everything is giftable and comes from a fair-trade community . it's great fun - but a bit pricey - and nothing here is a neccesity - but if you gotta buy gifts, at least here they are unique and helping another community . we love it! \\
\multicolumn{1}{c|}{} & \multicolumn{1}{c|}{} & Ours & my 8 middle old just loves it here - from musical injections to antiques to hand bags, everything is giftable and comes from a fair-trade community . it's great fun - but a bit pricey - and nothing here is a neccesity - but if you got rbis buy card, at least here they are unique and help another volunteering . we love it! \\ \hline \hline
\multicolumn{1}{c|}{\multirow{6}{*}{\rotatebox[origin=c]{90}{DP-BART}}} & \multicolumn{1}{c|}{\multirow{2}{*}{500}} & Naive & We have a great selection and a great one - we have a huge selection of a great quality and we have one of our favorite things! We have a free one of a while, and we were able to get a great deal and we had a really great selection. We were a huge one - and we got a really good deal on a great thing!  \\
\multicolumn{1}{c|}{} & \multicolumn{1}{c|}{} & Ours & OTHIcashDGDGcellaneous"startingtalking Cin win wi win in a cin in i win i w w win cin w w i wi in cin im w wi cin subtitles \\ \cline{2-4} 
\multicolumn{1}{c|}{} & \multicolumn{1}{c|}{\multirow{2}{*}{1000}} & Naive & My 8 year old just loves it here - it's a great place to shop for gifts and toys - and the kids love it here! - from the start of the year - we have a lot of fun with it! - and it is a bit pricey - but if you gotta buy something here, it's worth it.My 8 yr old just LOVES it here \\
\multicolumn{1}{c|}{} & \multicolumn{1}{c|}{} & Ours & Just a small-b\%\%\% -\%\% e-the-w\%\%-\%\% The e-all-s\%\% It's also a great place to be if you want to be a part of the community - but it's also very difficult to be in the community.It's a great complete \\ \cline{2-4} 
\multicolumn{1}{c|}{} & \multicolumn{1}{c|}{\multirow{2}{*}{1500}} & Naive & My 8 year old loves this store - it's a gift shop that has everything you need for your kids. The kids are all over the place. The store is small - but the kids love it.My 8 yr old just LOVES this store. It's a great gift shop. The prices are great - but if you gotta buy a gift, this is a great \\
\multicolumn{1}{c|}{} & \multicolumn{1}{c|}{} & Ours & This is the way we are going to go this year.We have a little bit of a way of doing this. The way we do it It's a bit pricey - but if you can't afford it, it's not a bad thing - and nothing here is cheap - but it's a little bit of a wonders
\end{tabular}
}
  \label{tab:examples}
\end{table*}

\end{document}